\documentclass[a4paper,12pt,fleqn]{article}
\usepackage{amsmath}

\newcommand{\pni}{\par\noindent}
\begin{document}
\title{Cosmology with two compactification scales}
\author{ A. G. Agnese\footnote{Email: agnese@ge.infn.it}\, and
M. La Camera\footnote{Email: lacamera@ge.infn.it}} 
\date{}
\maketitle 
\begin{center}
\em {Dipartimento di Fisica dell'Universit\`a di 
Genova\\Istituto Nazionale di Fisica Nucleare,Sezione di 
Genova\\Via Dodecaneso 33, 16146 Genova, Italy}\\
\end{center} 
\bigskip
\begin{abstract}\pni
We consider a $(4+d)$-dimensional spacetime broken up into a 
$(4-n)$-dimensional Minkowski spacetime (where $n$ goes from $1$ 
to $3$) and a compact $(n+d)$-dimensional manifold. 
At the present time the $n$ compactification radii are
of the order of the Universe size, while the other $d$ 
compactification radii are of the order of the Planck length. 
\end{abstract}
\bigskip \pni
\vspace{1in}\pni
\newpage
\baselineskip = 2\baselineskip

\section{Introduction} 

In recent years, there has been growing interest and a great deal 
of activity (see e.g. $[1],[2],[3],[4],[5]$) in multidimensional 
cosmology. A feature common to many of those works is to assume 
that the Universe is a $(4+d)$-dimensional manifold where, due to
its evolution, only three spatial dimensions are actually 
observable, while the remaining $d$ have curled up into compact  
spaces of unobservable small radii. This point of view is also 
apparent in the Kaluza-Klein spacetime of multidimensional 
supergravity$[6],[7],[8],[9]$.\pni
In this letter we consider dynamical compactification of a 
different sort. The $(4+d)$-dimensional space is  supposed to 
break up into a $(4-n)$-dimensional Minkowski space and a  
compact $(n+d)$-dimensional manifold whose compactification radii
are governed by Einstein's field equations. Here the integer $n$,
ranging from $1$ to $3$, is the number of usual spatial 
dimensions by hypothesis compactified, like the extra $d$ 
dimensions but with different radii, in a circle.  Moreover we 
require that at the initial time $t=0$ the compactification radii
of the $(n+d)$ spatial dimensions are all the same, and that each
radius of the $d$ extra dimensions at the present time  equals  
the Planck length, while each radius of the usual $n$ dimensions 
is comparable to the size of the Universe. In the simplified 
model we propose, only the cosmological constant $\Lambda$ will 
be retained in Einstein's equations, thus neglecting matter 
contributions as well as scalar field terms appropriate
to inflationary cosmology and a scale factor for the flat 
dimensions. This model is admittedly not realistic, but it can 
prove to be useful for future developments.\pni The letter is 
organised as follows: taking as guidelines the work of Chodos and
Detweiler $[10]$, we find the  solutions to the field 
equations and generalize those already found by Kasner 
when $\Lambda = 0$  $[11]$. Successively we consider  
eleven dimensional  cosmological models with different 
values of $n$ and $\Lambda$ and give numerical estimates of 
some quantities of interest such as the Universe age and the 
evolution of the compactification radii.\pni 

\section{The line element}

The metric suitable to our problem has the form
\begin{equation}
ds^{2} = dt^{2} - \sum_{\imath = 1}^{3-n}\, dx^{\imath}
dx^{\imath} - a^{2}(t)\, \sum_{\imath = 4-n}^{3}\, 
d\varphi^{\imath}d\varphi^{\imath} - l^{2}(t)\, 
\sum_{\imath = 1}^{d}\, d\psi^{\imath}d\psi^{\imath}
\end{equation}
where $a(t)$ and $l(t)$ are respectively the compactification 
radii  of each one of the $n$ and of the $d$ spatial dimensions 
and $\varphi^{\imath}$ and $\psi^{\imath}$ have a $2\pi$ period. 
\pni Einstein's equations, with cosmological term only, 
can be written as
\begin{equation}
R_{MN} = \dfrac{2 \Lambda}{n+d-1}\, g_{MN}\, , 
\hspace{1in}{\textsl {\scriptsize {M,N = 1,2,\ldots ,4+d}}}
\end{equation}
and the relevant ones are given explicitly by:
\begin{align*}
&n\, \dfrac{\ddot{a}}{a} + d\, \dfrac{\ddot{l}}{l} = \dfrac{2  
\Lambda}{n+d-1} \tag{3a}\\{}\\&\dfrac{\ddot{a}}{a} + 
(n-1)\,\left(\dfrac{\dot{a}}{a}\right)^{2} +d\, 
\dfrac{\dot{a}}{a}\, \dfrac{\dot{l}}{l} = \dfrac{2 
\Lambda}{n+d-1} \tag{3b}\\{}\\&\dfrac{\ddot{l}}{l} + 
(d-1)\,\left(\dfrac{\dot{l}}{l}\right)^{2} +n\, 
\dfrac{\dot{a}}{a}\, \frac{\dot{l}}{l}= \dfrac{2 \Lambda}{n+d
-1} \tag{3c}
\end{align*}
where a dot means derivative with respect to the time. \pni The 
system $(3)$ can be solved with the conditions that at the 
present time $t=t_{0}$ (age of the Universe) one 
has 
\setcounter{equation}{3}\begin{equation} 
\begin{split}a(t_{0}) = a_{0}\, , &\qquad\Dot{a}(t_{0})  = 
H_{0}\, a_{0} \\l(t_{0})  = l_{0}\, , &\qquad\Dot{l}(t_{0}) = 
h_{0}\, l_{0} \end{split}\end{equation}
The Hubble constant $H_{0}$ and the new constant $h_{0}$ which 
appear in the above conditions are not independent as one can see
from Eqs.(3) rewritten at $t=t_{0}$ with the introduction of the 
deceleration parameter $q_{0}= -\, (\Ddot{a} a/\Dot{a}^{2})_{0}$
and of its analogous $Q_{0}= -\, (\Ddot{l} l/\Dot{l}^{2})_{0}$: 
\begin{align*}
&n q_{0} H_{0}^{2} + d Q_{0} h_{0}^{2} = -\, \dfrac{2 \Lambda}
{n+d-1} \tag{5a} \\&(n-1-q_{0})\, H_{0}^{2} + d H_{0} h_{0} = 
\dfrac{2 \Lambda}{n+d-1} \tag{5b} \\&(d-1-Q_{0})\, h_{0}^{2} + 
n H_{0} h_{0} = \dfrac{2 \Lambda}{n+d-1} \tag{5c}  
\end{align*}
It is in fact straightforward to obtain:
\setcounter{equation}{5}\begin{equation}
\hspace{-0.4in} \dfrac{h_{0}}{H_{0}}=
\begin{cases}-\, \dfrac{n}{d-1} + \sqrt{\dfrac{n(n+d-1)} 
{d(d-1)^{2}} + \dfrac{2 \lambda}{d(d-1)}} &\qquad {\text {if}} 
\quad d \neq 1 \\ {} \\-\, \dfrac{n-1}{2} +\dfrac{\lambda}{n}\,  
&\qquad {\text {if}} \quad d = 1
\end{cases}\end{equation}
with 
\begin{equation}\lambda = 
\dfrac{\Lambda}{H_{0}^{2}}
\end{equation}
When $d\neq 1$ it must be $\lambda > - \, n(n+d-1)/2(d-1)$ for 
reality.\pni For future convenience we define the dimensionless 
quantities: 
\begin{align}
&\tau = H_{0}t \, , \qquad \tau_{0} = H_{0}t_{0} \\&\omega = 
\sqrt{\dfrac {n+d}{2\, (n+d-1)}\, |\lambda |} \\&\delta_{>} = 
\dfrac{1}{2}\, {\text {arctanh}}\, \dfrac{2 \omega}{n  + d\,  
\dfrac{h_{0}}{H_{0}}} \\&\delta_{<} = \dfrac{1}{2}\, {\text 
{arctan}}\, \dfrac{2 \omega}{n  + d\, \dfrac{h_{0}}{H_{0}}} 
\end{align}
The solutions to system $(3)$ can then be written as
\begin{equation}\hspace{-1in} 
\dfrac{a(\tau)}{a_{0}}=\begin{cases}&\left[ \dfrac{\sinh (\omega 
(\tau-\tau_{0}) + \delta_{>})} {\sinh 
\delta_{>}}\right]^{\beta_{+}}\, \left[ \dfrac{\cosh (\omega 
(\tau-\tau_{0}) + \delta
{>})} {\cosh \delta_{>}}\right]^{\beta_{-}} \hfill {\text {if}} 
\quad \lambda > 0 \\{}\\&\left[ 1 + (n  + d\, \dfrac{h_{0}}
{H_{0}})(\tau - \tau_{0}) \right]^{\beta_{+}} \hfill {\text 
{if}} \quad \lambda = 0 \\ {} \\&\left[ \dfrac{\sin (\omega 
(\tau-\tau_{0}) + \delta_{<})} {\sin \delta_{<}}\right]^
{\beta_{+}}\, \left[ \dfrac{\cos (\omega (\tau-\tau_{0}) + 
\delta_{<})} {\cos \delta_{<}}\right]^{\beta_{-}} \hfill 
{\text {if}} \quad \lambda < 0 
\end{cases}\end{equation}
\begin{equation}\hspace{-1in} 
\dfrac{l(\tau)}{l_{0}}=\begin{cases}&\left[ \dfrac{\sinh (\omega 
(\tau-\tau_{0}) + \delta_{>})} {\sinh \delta_{>}}\right]^
{\gamma_{-}}\, \left[ \dfrac{\cosh (\omega (\tau-\tau_{0}) + 
\delta_{>})} {\cosh \delta_{>}}\right]^{\gamma_{+}} \hfill 
{\text {if}} \quad \lambda > 0 \\{}\\&\left[ 1 + (n  + d\, 
\dfrac{h_{0}}{H_{0}})(\tau - \tau_{0}) \right]^{\gamma_{-}} 
\hfill {\text {if}} \quad \lambda = 0 \\ {} \\&\left[ \dfrac
{\sin (\omega (\tau-\tau_{0}) + \delta_{<})} {\sin \delta_{<}}
\right]^{\gamma_{-}}\, \left[ \dfrac{\cos (\omega (\tau-\tau_{0})
+ \delta_{<})} {\cos \delta_{<}}\right]^{\gamma_{+}} \hfill 
{\text {if}} \quad \lambda < 0 
\end{cases}\end{equation}
Here
\begin{align}
\beta_{+} & = \dfrac{1 
+ \sqrt {d\, (n+d-1)/n}}{n+d} \; ,&\beta_{-} & = \dfrac{1 - \sqrt
{d\, (n+d-1)/n}}{n+d} \\{}\nonumber \\\gamma_{+} & = \dfrac{1 + 
\sqrt {n\, (n+d-1)/d}}{n+d} \; ,&\gamma_{-} & = \dfrac{1 - \sqrt 
{n\, (n+d-1)/d}}{n+d} \end{align}
while $l_{0}$ is assumed to be of the order of the the Planck 
length and $a_{0}$ is the radius of the circle of the actually 
macroscopic dimensions.\pni Once $n$ and $d$ are fixed and 
$H_{0}$ is taken as known, the ratios $a(\tau)/a_{0}$ and 
$l(\tau)/l_{0}$ result to depend only on $\lambda$ and 
$\tau_{0}$. Then, due to the fact that these two quantities are 
not sufficiently well established, we might vary them step by 
step in a neighborhood, say, of $\lambda = 0$ and of $\tau_{0}=1$
to obtain numerical estimates of $a(\tau)/a_0$ and 
$l(\tau)/l_0$.\pni We find however more convenient to proceed in 
a different manner. Noticing that $\beta_{+} - \, \gamma_{-} = 
-\, (\beta_{-} - \, \gamma_{+}) \equiv 1/\alpha$ and defining 
\begin{equation}
\rho \equiv \left( \dfrac {l_{0}\, a(0)}{a_{0}\, l(0)} 
\right) ^{\alpha}
\end{equation}
which is expected to be a quantity much less than unity, one 
easily obtains from Equations $(12)$ and $(13)$ written at 
$\tau =0$:
\begin{equation}\hspace{-0.4in} 
\tau_{0} =\begin{cases}\dfrac{ \delta_{>} - {\text {arctanh}}
(\rho\, \tanh \delta_{>})}{\omega} &\qquad {\text {if}}\quad 
\lambda > 0 \\{}\\ \dfrac {1 -\, \rho}{n+d\, h_{0}/H_
{0}} &\qquad  {\text {if}}\quad \lambda = 0 \\{}\\ \dfrac{ 
\delta_{<} - {\text {arctan}}(\rho\, \tan \delta_{<})} {\omega} 
&\qquad {\text {if}}\quad \lambda < 0\end{cases}
\end{equation}
In this way we can calculate $\tau_{0}$ for a given $\lambda$ 
if we properly choose the parameter $\rho$ or, otherwise stated, 
the ratio $a(0)/l(0)$. To make an example we can recover, 
for $\lambda =0$, Kasner's solution by choosing $a(0)$ equals to 
zero, or equivalently  $l(0)$ equals to infinity, and therefore 
$\rho = 0$.\pni In view of the smallness of the ratio $l_0/a_0$, 
initial values for $a(0)$ and $l(0)$ of not too much different 
order of magnitude does not appreciably influence the results at 
farther times. Our choice is therefore to have at the initial 
time the same compactification radii for the $(n+d)$ spatial 
dimensions and so we put $a(0) = l(0)$. As a consequence, for a 
given pair $(n,d)$, we have only  $\lambda$ as the parameter left
free to evaluate both the actual age of the Universe $\tau_0$ and
the ratios $a(\tau )/a_0$ and $l(\tau )/l_0$.

\section{Numerical results for eleven dimensions}

We shall limit ourselves to the most popular choice of eleven 
dimensions and so fix the value $d = 7$. \pni
To begin with, let us examine the values of $\tau_0 = H_0 t_0$ 
which can be obtained by our model when the adimensional 
cosmological constant $\lambda$ vary in the interval $(-\, 1,1)$.
As to the Hubble constant $H_0$, it is common practice to write 
$H_0 = 100\eta\, km\, s^{-\, 1}\, Mpc^{-\, 1}$ where the 
uncertainty on it is put into the constant $\eta$, whose present 
value ranges from $0.50$ to $0.85$. Thus a characteristic time 
scale for the expansion of the Universe is the Hubble time $1/H_0
= (9.8/\eta)\, Gyr$.\pni
If we look at the graph of Figure $1$, where the above stated 
restriction on the negative values of $\lambda$ is apparent only 
for $n=1$, we can see that $\tau_0$ can exceed unity, as one 
expects, only if $n=1$ and $\lambda < 0$. Due however to the 
simplicity of our model, this fact does not seem a serious 
drawback and all other combinations of $n$ and $\lambda$ can not,
in our opinion, be ruled out. \pni
The sign of $\lambda$ and the values of $n$ appear to be of 
great importance for the time evolution of the radii $a(\tau )$ 
and $l(\tau )$ as it is shown in Figures from $2$~to~$7$. \pni 
We can summarize the various behaviours as follows: \pni
1) When $\lambda >0$, $a(\tau )$ is always increasing to 
infinity with time and so does $l(\tau )$ apart in the cases 
$n=2,3$ where $l(\tau )$ is initially decreasing in a finite time
interval. \pni 2) When $\lambda = 0$, $a(\tau )$ is always 
increasing to infinity with time, while $l(\tau )$ is constant if
$n=1$ and decreasing to zero if $n=2,3$. \pni
3) When $\lambda < 0$\,, $a(\tau )$ increases to infinity and 
$l(\tau )$ decreases to zero until $\tau$ reaches the finite 
value $\tau = \tau_0 + (\pi /2 - \delta_< )/\omega $\, ; whether 
this is a final state or a new initial state of the Universe is a
question we leave open. \pni
Let us notice, as one can see from Figures $3$,$5$ and $7$, that
the rate of variation of the Planck length is not so dramatic in 
the range of times considered, and in any case still compatible 
with the experimental bounds due to the possible time variation
of the fundamental constants involved in its definition.

\section{Conclusions}

The widespread belief in existing multidimensional cosmological 
models is that three spatial dimensions  expand isotropically 
while the remaining $d$ are actually curled up into spaces of 
dimensions comparable to the Planck length. Such a behaviour is 
exibited also by the $(4+d)$-dimensional Kaluza-Klein spacetime 
derived from M-theory. \pni
We instead propose that at least one of the three spatial 
macroscopic dimensions can undergo a compactification process 
with a consequent loss of isotropy. This fact would bring to 
important experimental consequences, for instance, with respect 
to the cosmic microwave background anisotropy. When all the three
usual spatial dimensions compactify, that space becomes like a 
flat three-dimensional one with the scale factor $a(\tau )/a_0$ 
describing its espansion. Of course in all the cases we have 
considered, the espansion in the distant future is driven by the 
cosmological costant.\pni Our model is admittedly greatly 
simplified, but it seems worth exploring the possibility of a 
compactification process also in the large scale.

\newpage
\begin{flushleft}
\Large\textbf {Figure captions}
\end{flushleft}
Figure 1: $\tau_0$ as a function of $\lambda$ when $n=1,2,3$ \pni
Figure 2: $\frac{a(\tau )}{a_0}$ as a function of $\tau$ when 
$n=1$     \pni
Figure 3: $\frac{l(\tau )}{l_0}$ as a function of $\tau$ when 
$n=1$     \pni
Figure 4: $\frac{a(\tau )}{a_0}$ as a function of $\tau$ when 
$n=2$     \pni
Figure 5: $\frac{l(\tau )}{l_0}$ as a function of $\tau$ when 
$n=2$     \pni
Figure 6: $\frac{a(\tau )}{a_0}$ as a function of $\tau$ when 
$n=3$     \pni
Figure 7: $\frac{l(\tau )}{l_0}$ as a function of $\tau$ when 
$n=3$ 
\newpage 
\begin{figure}
\begin{center}
\input{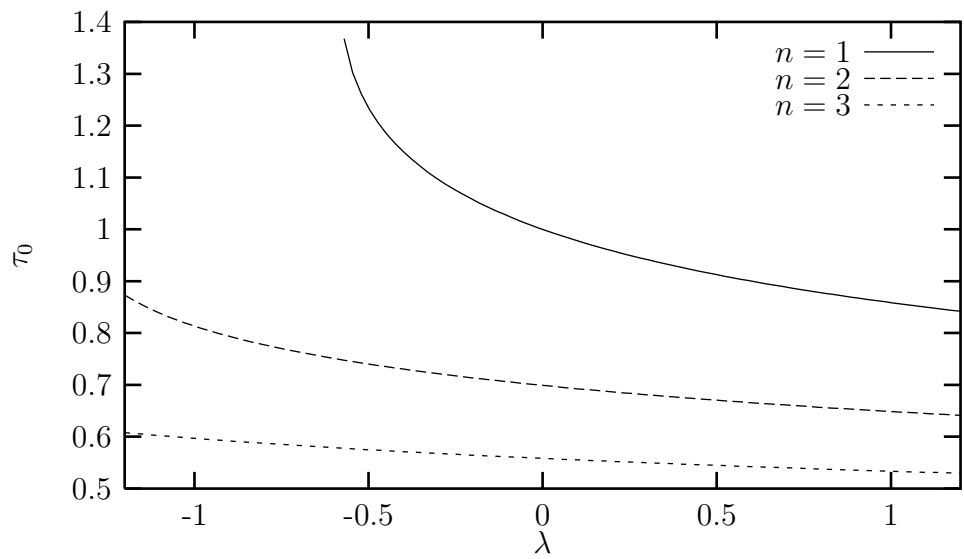}
\end{center}
\caption{$\tau_0$ as a function of $\lambda$ when $n=1,2,3$}
\end{figure}\hfil
\newpage
\begin{figure}
\begin{center}
\input{fig2}
\end{center}
\caption{$\frac{a(\tau )}{a_0}$ as a function of $\tau$ when 
$n=1$} 
\end{figure}
\begin{figure}
\begin{center}
\input{fig3}
\end{center}
\caption{$\frac{l(\tau )}{l_0}$ as a function of $\tau$ when 
$n=1$} 
\end{figure} 
\newpage
\begin{figure}
\begin{center}
\input{fig4}
\end{center}
\caption{$\frac{a(\tau )}{a_0}$ as a function of $\tau$ when 
$n=2$} 
\end{figure}
\begin{figure}
\begin{center}
\input{fig5}
\end{center}
\caption{$\frac{l(\tau )}{l_0}$ as a function of $\tau$ when 
$n=2$} 
\end{figure}
\newpage
\begin{figure}                 
\begin{center}
\input{fig6}
\end{center}
\caption{$\frac{a(\tau )}{a_0}$ as a function of $\tau$ when 
$n=3$} 
\end{figure}
\begin{figure}
\begin{center}
\input{fig7}
\end{center}
\caption{$\frac{l(\tau )}{l_0}$ as a function of $\tau$ when 
$n=3$}
\end{figure}

\end{document}